\documentstyle[psfig]{aipproc}

\begin{document}
\title{Closing the Low-mass Axigluon Window}

\author{Michael A. Doncheski}
\address{Department of Physics \\
Penn State University \\
Mont Alto, PA 17237}

\maketitle

\begin{abstract}
In this report, I will present the current status of the low-mass axigluon.  
The axigluon is a massive, color octet, axial vector boson, predicted in, 
{\it e.g.}, chiral color models and some technicolor models, with a mass of 
order the electroweak scale.  Axigluons with a mass larger than about 
$125 \; GeV$ to nearly $1 \; TeV$ can be eliminated by di-jet production at 
hadron colliders like the TEVATRON, but a low-mass window exists that the 
di-jet search can not probe.  $\Upsilon$ decays can rule out axigluons with a 
mass up to $25 \; GeV$, and low energy $e^+ e^-$ (PEP and PETRA) can rule out 
axigluons with a mass up to $50 \; GeV$ using a measurement of $R$.  Top 
production at the TEVATRON disfavors a light axigluon.  A measurement of $R$ 
at LEP {\bf strongly} disfavors a light axigluon, and rules out an axigluon 
with mass $< \; 365 \; GeV$.
\end{abstract}

\section*{Motivation}

The possible existence of an axigluon was first realized in chiral color 
models~\cite{chiral}, where the gauge group of the strong interaction is 
extended from $SU(3)_C$ to $SU(3)_L \times SU(3)_R$.  At low energy, this 
larger color gauge group breaks to the usual $SU(3)_C$ with its octet of 
massless vector gluons, but it leaves a residual $SU(3)$ with an octet of 
massive, axial vector bosons called {\it axigluons}.  In these chiral color 
models, the axigluon is expected to have a mass of order the electroweak 
scale.  Similar states are predicted in technicolor models~\cite{tc}.

In order to search for these states, Bagger, Schmidt and King~\cite{bsk} noted 
that the di-jet cross section at hadron colliders would be modified by the 
addition of $s$-channel axigluon exchange.  Searches were performed by the UA1 
and CDF collaborations, with limits of $150 \; GeV  <  M_A  < 310 \; GeV$ by 
UA1~\cite{UA1-dijet} and $120 \; GeV  <  M_A  < 980 \; GeV$ by 
CDF~\cite{CDF-dijet}.  Given additional center of mass energy and/or 
luminosity, these di-jet searched at hadron colliders will easily raise the 
upper exclusion limit, but it will be difficult to decrease the lower 
exclusion limit.

Several additional search strategies were suggested involving the $Z^0$ and 
the large amounts of data taken by the LEP experiments.  Rizzo~\cite{rizzo} 
suggested $Z^0 \rightarrow q \bar{q} A$ and Carlson, 
{\it et al.},~\cite{carlson} suggested $Z^0 \rightarrow g A$ going through a 
quark loop.  These suggestions involve low rates, and the former requires 
precision multi-jet reconstruction.

In the remainder of this report, I will address some additional search 
strategies for the axigluon, and report on the current status of the search.

\section*{$\Upsilon$ decays to real axigluons}

The decay of the $\Upsilon$ family is an ideal area to search for low mass, 
strongly interacting particles.  In the Standard Model, the dominant hadronic 
decay mode of any heavy vector quarkonium state ($J^{PC} = 1^{--}$) is the 3 
gluon mode, $V_Q \rightarrow ggg$, where $Q$ refers to the specific flavor of 
heavy quark.  The decay to a single gluon is forbidden by color, while the 
decay to 2 gluons is forbidden by both the Landau-Pomeranchuk-Yang 
theorem~\cite{lpy} (which forbids the decay of a $J=1$ state to 2 massless 
spin 1 states) and quantum numbers ($C = -1$ for the gluon, so an odd number 
of gluons are needed for this particular decay).  The leading order decay rate 
of a heavy quarkonium state to 3 gluons is well known:
\begin{equation}
\Gamma(V_Q \rightarrow ggg) = \frac{40 (\pi^2 - 9) \alpha_s^3}{81 \pi M_V^2} 
| R(0) |^2
\end{equation}
where $M_V$ is the quarkonium state's mass and R(0) is the non-relativistic, 
radial wavefunction evaluated at the origin.

A heavy, vector quarkonium state {\bf may} decay into a gluon plus an 
axigluon.  As the axigluon is massive, the Landau-Pomeranchuk-Yang theorem is 
avoided, and the axigluon has $C = +1$.  The decay rate for 
$V_Q \rightarrow Ag$ is given by~\cite{mhr1}:
\begin{equation}
\Gamma(V_Q \rightarrow Ag) = \frac{16 \alpha_s^2}{9 M_V^2} | R(0) |^2 
(1 - x) (1 + \frac{1}{x})
\end{equation}
where $x = \left( \frac{M_A}{M_V} \right)^2$.  Both this decay rate and the 
leading order Standard Model rate depend on the non-relativistic radial 
wavefunction; a ratio of these two decay rates does not depend on the 
wavefunction, and, as such, had much less uncertainty.  The ratio is given by:
\begin{equation}
\frac{\Gamma(V_Q \rightarrow Ag)}{\Gamma(V_Q \rightarrow ggg)} = 
\frac{18 \pi}{5 \alpha_s (\pi^2 - 9)} (1 - x) (1 + \frac{1}{x})
\end{equation}
Notice that, since the gluon plus axigluon mode has one fewer power of 
$\alpha_s$, the ratio is large (the numerical factor in front of the 
kinematical structure is approximately 100).

\begin{figure}[ht] 
\centerline{\psfig{file=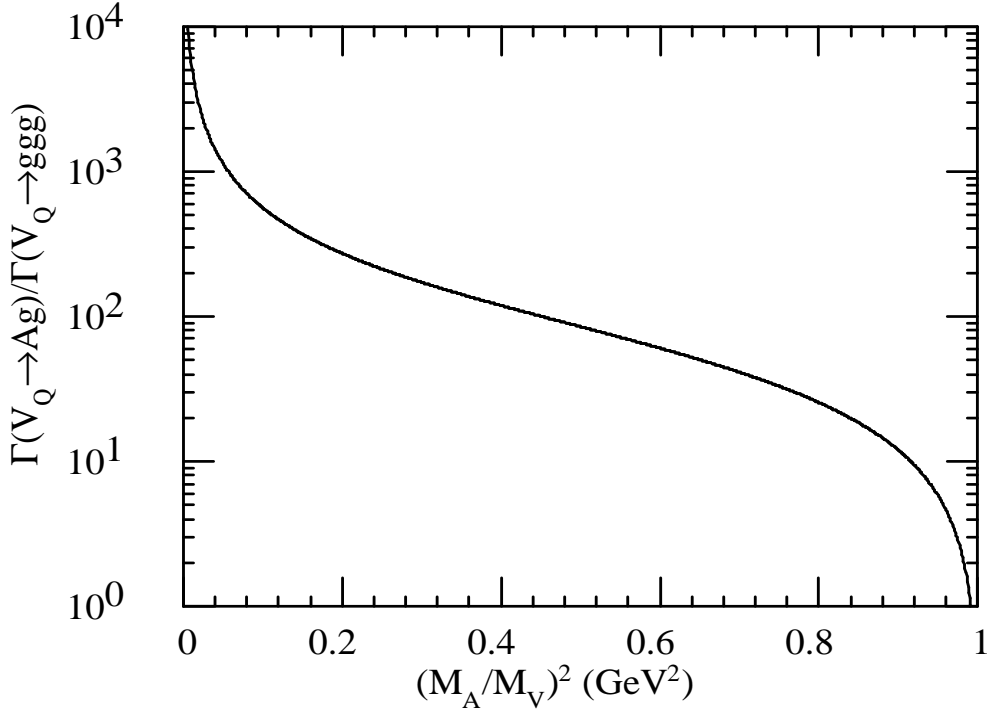,height=10.625cm,width=14.375cm,angle=-90}}
\vspace{10pt}
\caption{Ratio $\frac{\mbox{$\Gamma(V_Q \rightarrow Ag)$}}
{\mbox{$\Gamma(V_Q \rightarrow ggg)$}}$ for the decay of a heavy quarkonium 
state to a real axigluon.}
\label{fig1}
\end{figure}

This ratio, as a function of $x$ is shown in Figure 1.  The addition of this 
new hadronic decay mode will at least double the hadronic width of a vector 
quarkonium state, even for an axigluon mass nearly equal to the quarkonium 
state mass.  Using this process and the $\Upsilon$ system, we can exclude an 
axigluon with mass below about $10 \; GeV$.  A analysis by Cuypers and 
Frampton~\cite{cf} yielded quantitatively similar conclusions.

\section*{$\Upsilon$ decays to virtual axigluons}

In addition to $\Upsilon$ decays to real axigluons, it is possible to study 
$\Upsilon$ decays to virtual axigluons, 
$\Upsilon \rightarrow g A^* (\rightarrow q \bar{q})$.  The decay rate is given 
by~\cite{mhr2}
\begin{equation}
\Gamma(V_Q \rightarrow q \bar{q} g) = \frac{2^8 n  \alpha_s^3}{3^5 \pi} 
\frac{M_V^2}{M_A^4} F(x) | R(0) |^2 
\end{equation}
where $n$ is the number of active quark flavors (in this case 4) and
\begin{equation}
F(x) = \frac{3}{2} x^2 \left( 2 x \ln \left( \frac{x}{x - 1} \right) - 2 - 
\frac{1}{x} 
\right).
\end{equation}
As before, we can look at the ratio of this hadronic width to the dominant 
Standard Model width:
\begin{equation}
\frac{\Gamma(V_Q \rightarrow q \bar{q} g)}{\Gamma(V_Q \rightarrow ggg)} = 
\frac{128 F(x)}{15 (\pi^2 - 9) x^2}.
\end{equation}
This time, there is no large numerical factor.  This ratio is shown in 
Figure~2.  The dashed lines in the figure indicate 2 possible exclusion limits 
that can be made using data.  The more conservative estimate is to argue that 
our knowledge of the $\Upsilon$ width is such that a correction to the 
standard width larger than 50\% is unacceptable; thus, this ratio is smaller 
than 0.5, which gives an upper exclusion limit of $M_A < 21 \; GeV$.  A less 
conservative estimate is to compare this correction to the expected rate to 
QCD radiative corrections to the Standard Model rate and other possible 
contributions to the hadronic width ({\it e.g.,} 
$\Upsilon \rightarrow \gamma^* \rightarrow q \bar{q}$), and argue that another 
correction larger than these is unacceptable.  In this case, the ratio must be 
less than 0.25, excluding axigluons with mass smaller than $25 \; GeV$.

\begin{figure}[ht] 
\centerline{\psfig{file=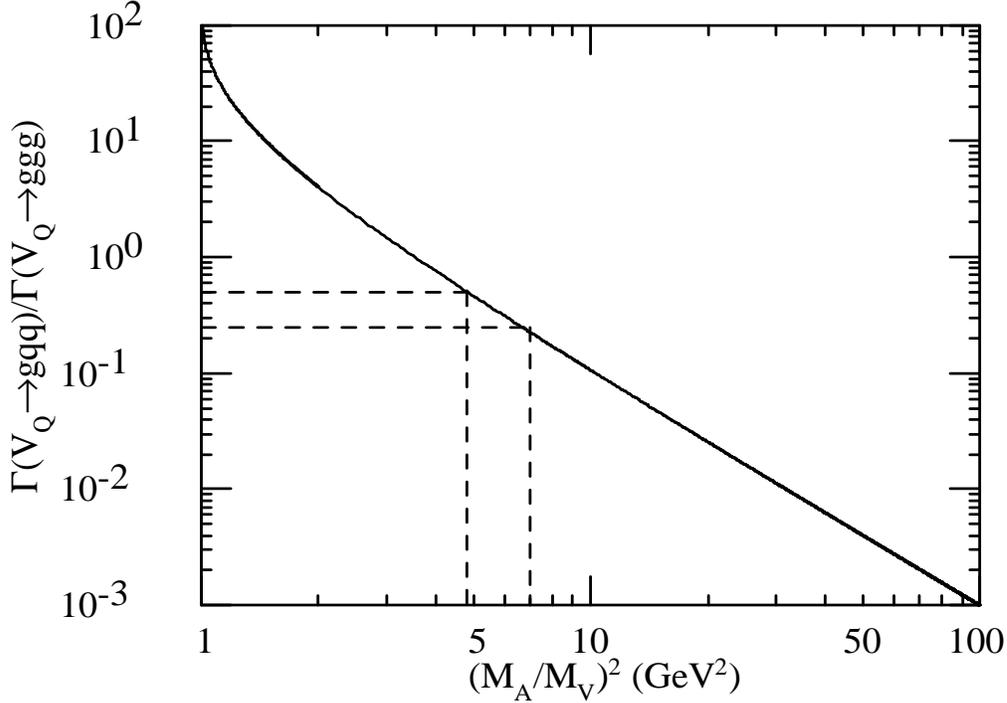,height=10.625cm,width=14.375cm,angle=-90}}
\vspace{10pt}
\caption{Ratio $\frac{\mbox{$\Gamma(V_Q \rightarrow q \bar{q} A)$}}
{\mbox{$\Gamma(V_Q \rightarrow ggg)$}}$ for the decay of a heavy quarkonium 
state to a virtual axigluon.  The dashed lines indicate 2 possible exclusion 
limits.}
\label{fig2}
\end{figure}

Not long after our work on the $\Upsilon$, Cuypers and Frampton and Cuypers, 
Falk and Frampton~\cite{cff} published papers on the $R$ value in $e^+ e^-$ 
collisions at low energy.  They included the full set of QCD radiative 
corrections, including axigluon radiative corrections, to the tree level 
process.  They exclude an axigluon with $M_A < 50 \; GeV$ using PEP and PETRA 
data.

\section*{Top production}

The top is too short lived to allow for a toponium state; if it did, the same 
techniques that worked in the $\Upsilon$ system would work for toponium as 
well.  On the other hand, because of the large mass of the top, top production 
is inherently perturbative, $q \bar{q} \rightarrow t \bar{t}$ is well 
understood, and it can be used to search for a light axigluon.  The parton 
level cross section for $q \bar{q} \rightarrow t \bar{t}$, due to an 
$s$-channel gluon, is well known:
\begin{equation}
\left( \frac{d \sigma}{d \hat{t}} \right)_0 = 
\frac{1}{16 \pi \hat{s}^2} \frac{64 \pi^2}{9} \alpha_s^2 
\left[ \frac{ (m^2 - \hat{t})^2 + (m^2 - \hat{u})^2 + 2 m^2 \hat{s}}{\hat{s}^2}
\right]
\end{equation}
and the cross section with the addition of an $s$-channel axigluon 
is~\cite{mr1}
\begin{equation}
\left( \frac{d \sigma}{d \hat{t}} \right)_{q \bar{q}} = 
\left( \frac{d \sigma}{d \hat{t}} \right)_0 
\left[ 1 + |r(\hat{s})|^2 + 4 \Re(r(\hat{s})) 
\frac{(\hat{t} - \hat{u}) \hat{s} \beta}{(\hat{t} - \hat{u})^2 
+ \hat{s}^2 \beta^2} \right]
\end{equation}
where $r(\hat{s}) = \frac{\mbox{$\hat{s}$}}
{\mbox{$\hat{s} - M_A^2 + i M_A \Gamma_A$}}$ and $\beta$ is the top quark 
velocity parameter, 
$\beta = \sqrt{1 - \frac{\mbox{$4 m^2$}}{\mbox{$\hat{s}$}}}$.  The addition of 
an $s$-channel axigluon affects both the total cross section and the 
forward-backward asymmetry (only the interference term affects the 
forward-backward asymmetry).

The results on total cross section are shown in Figure~3.  From the relatively 
good agreement between experimental values of the top cross 
section~\cite{top-D0,top-CDF} and theoretical calculations~\cite{bc,cat}, we 
can say that an axigluon is disfavored by top production cross section, but 
nothing conclusive can be said.

\begin{figure}[ht] 
\centerline{\psfig{file=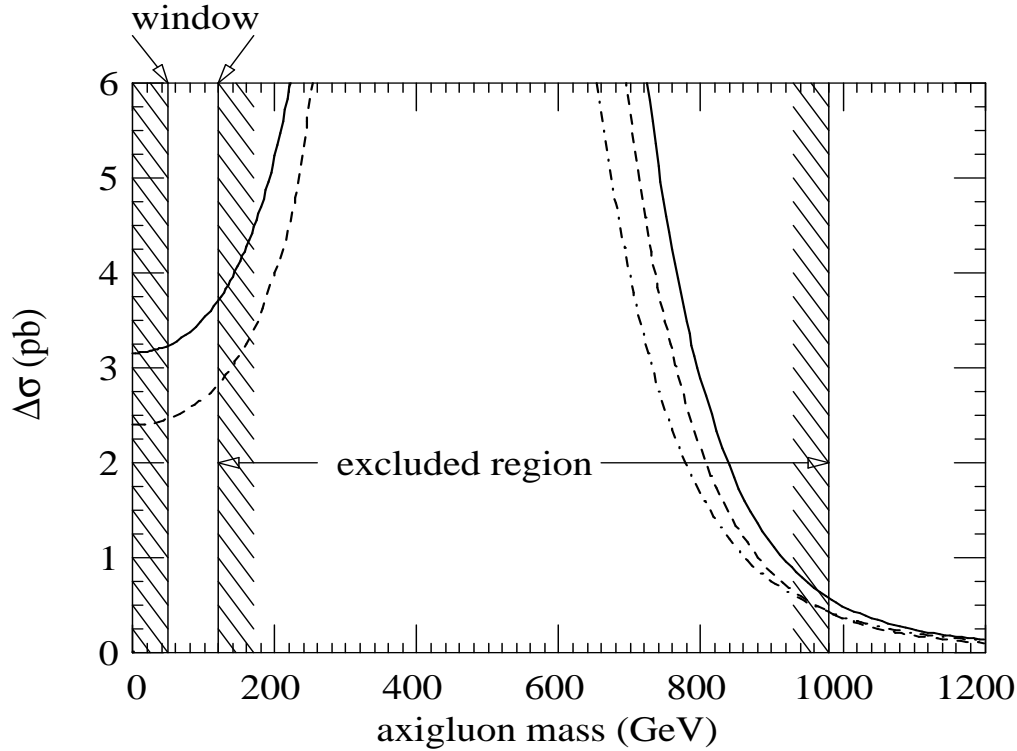,height=10.625cm,width=14.375cm,angle=-90}}
\vspace{10pt}
\caption{Difference in top production cross section, based on the presence 
of an axigluon.  The different curves are for different sets of leading order 
parton distribution functions, and different choices for axigluon width.  
The solid (dotdashed) line is for the ``new'' Duke and Owens pdf's with 
$\Gamma_A = 0.1 M_A$ ($0.2 M_A$); the dashed line is for CTEQ4L with 
$\Gamma_A = 0.1 M_A$.}
\label{fig3}
\end{figure}

\begin{figure}[ht] 
\centerline{\psfig{file=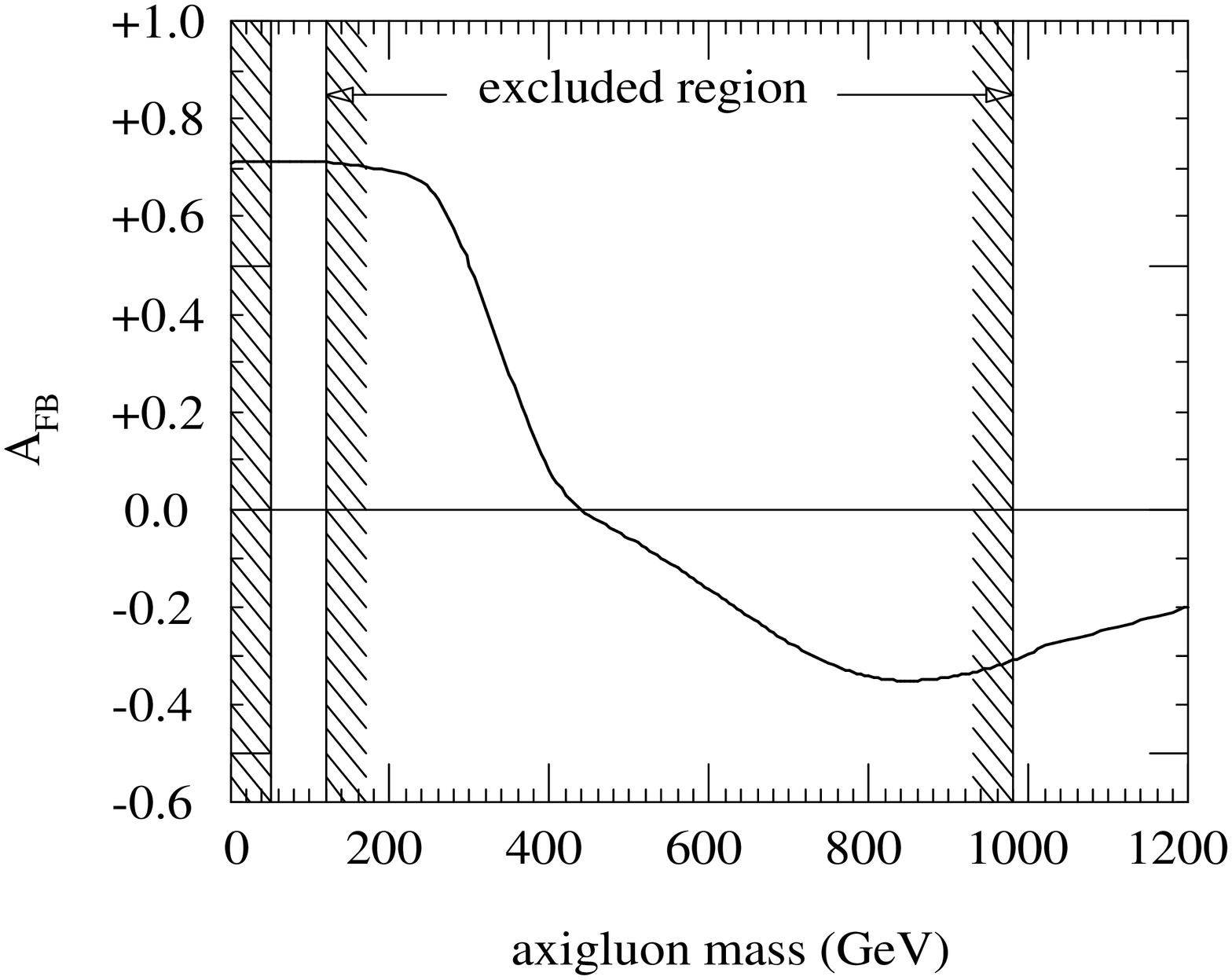,height=10.625cm,width=14.375cm,angle=-90}}
\vspace{10pt}
\caption{Top production forward-backward asymmetry.}
\label{fig4}
\end{figure}

Shown in Figure~4 is the forward-backward asymmetry in top production as a 
function of axigluon mass.  Without an axigluon, the asymmetry is identically 
zero.

\section*{Miscellaneous}

Unitarity is violated, in that $Q \bar{Q} \rightarrow Q \bar{Q}$ will be 
non-perturbative unless
\begin{equation}
M_A > \sqrt{\frac{5 \alpha_s}{3}} M_Q
\end{equation}
as pointed out by Robinett~\cite{rick}.  Using the top quark as $Q$, this 
leads to a lower limit on the axigluon mass of $M_A > 73\;GeV$~\cite{mr2}.

Higgs searches, {\it e.g.}, by CDF, can make use of the process 
$p \bar{p} \rightarrow W + X^0$, where $X^0$ is the neutral Higgs boson, and 
it is assumed to decay to $b \bar{b}$~\cite{bhat}.  The limit on Higgs boson 
mass is such that $\sigma \cdot BR > 15-20 \; pb$ are not allowed.  The same 
final state is possible with an axigluon in place of the Higgs boson; we find 
the part level cross section for $q \bar{q}' \rightarrow W A$ to be~\cite{mr2}:
\begin{equation}
\frac{d \hat{\sigma}}{d \hat{t}} = 
\frac{4 \alpha_s}{9} \left[ \frac{G_F M_W^2}{\sqrt{2}} \right ]
\frac{|V_{qq'}|^2}{\hat{u} \hat{t} \hat{s}^2} 
\left[ \hat{u}^2 + \hat{t}^2 + 2 \hat{s} (M_W^2 + M_A^2) - 
\frac{M_A^2 M_W^2 (\hat{u}^2 + \hat{t}^2)}{\hat{u} \hat{t}} \right].
\end{equation}
Assuming $BR(A \rightarrow b \bar{b}) = \frac{1}{5}$, and calculating the 
cross section for the associated production of $W + A$, a conservative lower 
limit of $M_A > 70\;GeV$ is possible, using the same analysis at the Higgs 
search.  

Finally, we can examine the value of $\alpha_s$, extracted from low energy 
data but run up to $M_Z$ to the value of $\alpha_s$ extracted from the 
hadronic width of the $Z^0$ at the pole.  Since the axigluon mass is expected 
to be at least $70\;GeV$, the running of $\alpha_s$ should not be affected 
much by the axigluon.  Then, the $R$ value at low energy, or the hadronic 
width at the $Z^0$ pole, is subject to a correction from real and virtual 
axigluons~\cite{cff}, of the form:
\begin{equation}
\left[ 1 + \frac{\alpha_s(\sqrt{s})}{\pi} 
f \left( \frac{\sqrt{s}}{M_A} \right) + {\cal O}(\alpha_s^2) \right]
\end{equation}
where the function $F(\sqrt{s}/M_A)$ is calculated in Ref.~\cite{cff}.  The 
Particle Data Group~\cite{pdg} quotes a value of $\alpha_s$ from various low 
energy data run up to $M_Z$ as $\alpha_s^{(LE)} = 0.118 \pm 0.004$, while the 
extraction from the hadronic with of the $Z^0$ at $M_Z$ as 
$\alpha_s^{(HE)} = 0.123 \pm 0.004 \pm 0.002$.  Attributing the difference in 
the extracted values of $\alpha_s$ to the axigluon gives abound on the 
$f(\sqrt{s}/M_A)$ term, such that $f(M_Z/M_A) \leq 0.042 \pm 0.050$.  This 
implies that $f(M_Z/M_A) < 0.092 (0.142)$ at the $65\%$ $(95\%)$ level, and 
that $M_Z > 570\;GeV (365\;GeV)$ at the same confidence levels.  Should the 
agreement between the low energy and high energy extractions of $\alpha_s$ 
increase, the corresponding lower limit on the axigluon mass would also 
increase.

\section*{Conclusions}
 
The existence of an axigluon is predicted in chiral color models.  A low-mass 
axigluon is difficult to exclude in typical collider experiments ({\it e.g.}, 
using di-jet data).  Other approaches must be used to rule out axigluons with 
masses below $125\;GeV$.  $\Upsilon$ decay, top production, unitarity bounds 
and associated production of a $W$ boson with an axigluon can exclude 
axigluons with mass below about $70\;GeV$.  A comparison of $\alpha_s$ as 
extracted in low energy experiments and high energy experiments can rule out 
an axigluon with a mass lower than $365\;GeV$.  This completely closes the 
low-mass axigluon window, and when combined with the CDF limits, an axigluon 
with mass below about $1\;TeV$ is not allowed.  An axigluon, if it exists, is 
in the realm of $TeV$ physics.

\section*{acknowledgements}
I would like to acknowledge the support of Penn State through a Research 
Development Grant and the Mont Alto Faculty Affairs Committee's Professional 
Development Fund.  I would like to thank Rick Robinett for carefully reading 
this manuscript.

\end{document}